\documentclass[epj]{svjour}
\usepackage{graphicx}
\usepackage{amsmath}
\usepackage{siunitx}

\newcommand{\mat}[1]{\mathbf{#1}}
\DeclareMathOperator{\tr}{tr}
\DeclareMathOperator{\var}{var}
\newcommand{\frobnorm}[1]{\left\| #1 \right\|_\mathrm{F} }

\usepackage[usenames,dvipsnames]{xcolor}

\begin{document}
\title{Analysis of localization-delocalization transitions in corner-sharing tetrahedral lattices}
\titlerunning{LD transition in CST lattices}
\author{Martin Puschmann, Philipp Cain, \and Michael Schreiber}         
\authorrunning{Martin Puschmann \textit{et al.}}
\institute{Institute of Physics, Chemnitz University of Technology, D-09107 Chemnitz, Germany}
\date{Received:  / Revised version: }
\abstract{
We study the critical behavior of the Anderson localization-delocalization transition in corner-sharing tetrahedral lattices. We compare our results obtained by three different numerical methods namely the multifractal analysis, the Green resolvent method, and the energy-level statistics which yield the singularity strength, the decay length of the wave functions, and the (integrated) energy-level distribution, respectively. From these measures a finite-size scaling approach allows us to determine the critical parameters simultaneously. With particular emphasis we calculate the propagation of the statistical errors by a Monte-Carlo method. We find a high agreement between the results of all methods and we can estimate the highest critical disorder $W_\mathrm{c}=14.474(8)$ at energy $E_\mathrm{c}=-4.0$ and the critical exponent $\nu=1.565(11)$. Our results agree with a previous study by Fazileh {\it et al.} but improve accuracy significantly.
}
\maketitle

\section{Introduction}\label{sec:intro}

On a very fundamental level the Anderson model of localization \cite{And58} demonstrates the influence of quantum interference on the diffusion of an electron within a lattice. The variation of disorder, e.g. by a specific choice of random on-site potential energies, leads to a phase transition of the electronic wave functions between localization and delocalization (LD) corresponding to insulating and metallic behavior, respectively. The critical behavior at the LD transition is described quantitatively by the scaling theory of localization \cite{Weg76,AbrALR79}. In particular, the localization length and the conductance behave according to a power law in the vicinity of the transition. The corresponding critical exponents are expected to adopt universal values as predicted by random matrix theory \cite{Wig51,Dys62} meaning their values depend only on the classification according to the symmetry of the underlying Hamiltonian. Universal results obtained for the rather abstract Anderson model of localization are therefore applicable also to complex experimental measurements \cite{LopCSGD12}.  

The Anderson LD transition was studied extensively during the last decades (for reviews see \cite{KraM93,BraK03,EveM08}). 
A major subject of theoretical research covered the numerical verification of the concept of universality.
Here considerable progress was achieved during the last decade due to high accuracy calculations combined with extensive data and error analysis \cite{RodVSR11,SleO14,UjfV15}.
 
In this work we study the Anderson LD transition in corner-sharing tetrahedral (CST) lattices. The CST lattice appears in normal spinel structures as a non-bipartite sublattice, e.g. the $\rm Ti$ atoms in a ${\rm Li Ti_2 O_4}$ crystal. Gradual replacement of $\rm Ti $ atoms by $\rm Al $ atoms leads to a transition from metallic to insulating behavior. Recent numerical analyses \cite{FazGJ04,FazCGT06} of this LD transition focused on an estimate for the critical concentration of $\rm Al$. The results were inconsistent with experiments which was attributed to the use of one-electron models missing electron-electron correlations \cite{FazGJ04,FazCGT06}. 
Nevertheless the LD transition in a CST lattice should obey the predictions of random matrix theory. According to the symmetry of the corresponding Anderson Hamiltonian the model belongs to the Gaussian orthogonal ensemble (GOE). For GOE in three dimensions the most accurate estimate of the critical exponent $\nu$ of the localization length \cite{SleO14} is obtained for a simple cubic (sc) lattice by the transfer-matrix method \cite{PicS81a}. The result of $\nu=1.571\pm0.008$ agrees well with other recent results from multifractal analyses of the wave functions \cite{RodVSR11,VasRR08,RodVR08b}. Since the symmetry of the Hamiltonian is independent of the crystal lattice itself, the same universal exponent $\nu$ is expected also for other lattice geometries, which was demonstrated for the face-centered-cubic (fcc) and the body-centered-cubic (bcc) lattice \cite{EilFR08}. 
In this paper we investigate the critical behavior, i.e. the exponent $\nu$ of the localization length, for the LD transition in a CST lattice. We compare results of three different methods. Primarily we use the multifractal analysis (MFA) to obtain the phase diagram. This method utilizes multifractal fluctuations of the wave functions at the critical point. Secondarily we apply the Green-resolvent method (GRM) to calculate the exponential decay of a wave function along a bar-shape geometry. The commonly preferred transfer-matrix method (TMM) is not applicable in our case because of the connectivity\footnote{For all crystal orientations the construction leads to singular link matrices between successive layers. Also, a subdivision of a layer into several sublayers as it was done in triangular lattices~\cite{SchO91} would not help.} of the CST lattice. As a third independent approach we study the distribution of energy-level spacings (ELS). Because of the finite system size of the calculations all three methods require a finite-size-scaling (FSS) procedure in order to obtain the critical parameters.

This paper is organized as follows. In Sec.~\ref{sec:Anderson}, we shortly introduce the Anderson model of localization and describe the Hamiltonian for the CST lattice. In the following Section we explain the MFA, GRM, ELS, and the FSS approach. In Sec.~\ref{sec:Results} we present our numerical results and compare the methods. Finally, in Sec.~\ref{sec:Conclusion} we conclude with a short summary.

\section{Anderson model of localization and the corner-sharing tetrahedral lattices}\label{sec:Anderson}
\begin{figure}
\centering
\includegraphics[]{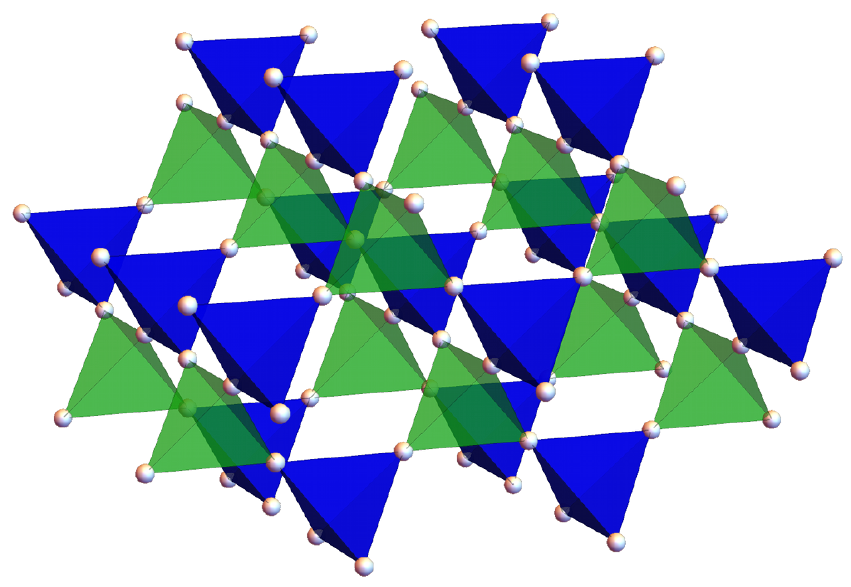}
\caption{A part of the CST lattice. The spheres embody the lattice sites and the edges of the tetrahedrons epitomize the nearest neighbor connections. The coloring of the tetrahedrons indicates the two translation invariant groups of tetrahedrons. }
\label{fig:CSTL_PART}
\end{figure}

We consider the Anderson Hamiltonian for non-interacting electrons in site representation 
\begin{align}
\label{eqn:AH}
H=\sum\limits_{i}\upsilon_i\left|i\right\rangle \left\langle i\right|-\sum\limits_{i,j}^\mathrm{n.n.}\left|i\right\rangle \left\langle j\right|\quad.
\end{align}
$\upsilon_i$ are randomly chosen on-site potentials distributed uniformly in the interval $\left[-W/2,W/2\right]$. The interval width $W$ parametrizes the disorder strength. Every site is connected to its six nearest neighbor sites according to the structure of the CST lattice as shown in Fig.~\ref{fig:CSTL_PART}.

\begin{figure}
\centering
\includegraphics[]{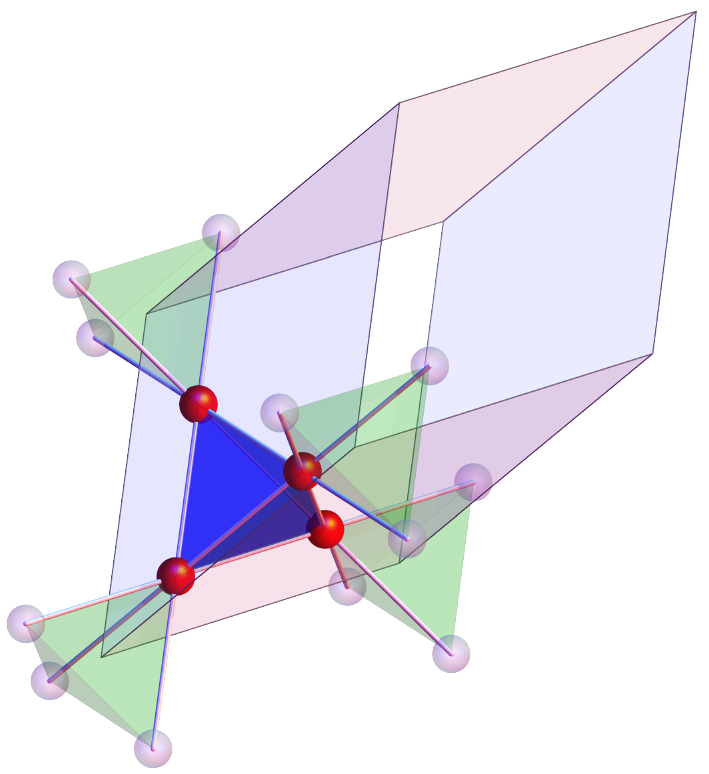}
\caption{Primitive UC of the CST lattice. The shaded rhombohedron represents the Bravais lattice, the red spheres the basis sites. The gray spheres are sites of neighboring UCs. On each corner of the rhombohedron is one green tetrahedron (not all shown). Only the visualized green tetrahedrons have a connection with the blue tetrahedron in this UC.}
\label{fig:CSTL_pUC}
\end{figure}

The CST lattice is built from regular tetrahedrons. The shared corners represent the lattice sites and the edges the nearest neighbor connections. The different colors distinguish tetrahedrons that can be transformed into each other by simple translations. The CST lattice is a member of the fcc Bravais-lattice class. We build up the lattice using primitive unit cells (UCs) as the smallest possible building block as illustrated in Fig.~\ref{fig:CSTL_pUC}. It contains four lattice sites. The six edges of the blue tetrahedron represent the intra UC connections. Taking into account the translational symmetry, there are also six different inter UC connections visualized by the edges of the green tetrahedrons. The three Bravais-lattice vectors span a rhombohedron. Because the CST lattice is a sublattice only, half of the rhombohedron is empty. Note that alternatively the CST lattice can be constructed also in a non-primitive way from a fcc lattice. Such a UC contains 16 sites. In the following we will express the system size always in units of primitive UCs stacked along the three Bravais-lattice vectors. In general we use periodic boundary conditions. 

\section{Methods}\label{sec:Methods}
\subsection{Finite-size-scaling approach}\label{sec:FSS}
	The LD transition is a second-order phase transition characterized by the algebraic divergence of the correlation length $\xi(\omega)\propto|\omega|^{-\nu}$ as the transition is approached ($\omega\rightarrow 0$). The dimensionless scaling variable $\omega= \frac{\rho-\rho_c}{\rho_c}$ defines the distance from the critical point $\rho_\mathrm{c}$, in our case in terms of either energy ($\rho=E$) or disorder strength ($\rho=W$).
	This behavior is observed only for an infinite system while our calculations are limited to finite system sizes. We therefore employ a FSS ansatz in order to extract the critical parameters of the transition. The implementation is based on an expansion function for the measured quantity close to the critical point and considers both relevant and irrelevant influences \cite{SleO99a}. Irrelevant contributions are quantified by an irrelevant exponent $y$ describing finite-size corrections that eventually disappear with increasing system size.
	Let us consider e.g. the localization length $\Lambda(\omega,L)$ for a system with characteristic length $L$. Nonlinearities are taken into account by expanding the $\omega$~dependence in terms of Taylor polynomials and the dependence on $L$ in terms of Chebyshev polynomials of the first kind $T_k(\cdot)$. For the expansion orders $n_\mathrm{r}$, $m_\mathrm{i}$, $n_\mathrm{r}$, and $m_\mathrm{i}$  the scaling function is given by
	\begin{align}
	\label{eq:fss}
	\Lambda=F\left[\Omega_\mathrm{r}L^\frac{1}{\nu},\Omega_\mathrm{i}L^{-y}\right]=\sum\limits_{n=0}^{n_\mathrm{i}}\frac{\Omega_\mathrm{i}^nL^{-ny}}{n!}F_n\left[\Omega_\mathrm{r}L^\frac{1}{\nu}\right]
	\end{align}
	with
	\begin{align}
	F_n\left[\Omega_\mathrm{r}L^\frac{1}{\nu}\right]&=\sum\limits_{k=0}^{n_\mathrm{r}}a_{n,k}T_k(\Omega_\mathrm{r}L^\frac{1}{\nu})\quad,\\
	\Omega_\mathrm{r}&=\omega+\sum\limits_{n=2}^{m_\mathrm{r}}\frac{b_n}{n!}\omega^n\quad\text{, and}\\
	\Omega_\mathrm{i}&=1+\sum\limits_{n=1}^{m_\mathrm{i}}\frac{c_n}{n!}\omega^n\quad.
	\end{align}
	This expansion contains $N_\mathrm{P}=(n_\mathrm{i}+1)(n_\mathrm{r}+1)+m_\mathrm{i}+m_\mathrm{r} +1+ \Theta(n_\mathrm{i})$ free parameters to be determined by a nonlinear fitting procedure.  

	In order to estimate the errors associated with the nonlinear fit reliably, we use a Monte-Carlo method \cite{RodVSR11}. We generate at least $10^4$ synthetic datasets from our original data by adding Gaussian noise to each data point. The individual error of each original data point defines the standard deviation of the noise. Combining the fit results for all synthetic datasets we obtain distributions for the critical parameters. 
	In general these distributions have a Gaussian shape which allows us to identify standard deviations easily. In contrast to this stochastic inaccuracy it is more complicated to identify systematic deviations. We use several expansion orders to reduce the influences of a specific fitting function. In some cases we observe deviations from Gaussian distributions, which we attribute to local minima of the $\chi^2$ sum in the nonlinear fit. 
	Their occurrence can be reduced by a reasonable choice of smaller expansion orders.

	\subsection{Multifractal analysis}\label{sec:MFA}
	In the vicinity of the LD transition multifractal fluctuations of the wave functions are observed~\cite{CasP86,SchG91}. The MFA quantifies these fluctuations by a box-counting algorithm using different moments of the probability distribution.  
	
	A system of $L^3$ UCs is subdivided into boxes of size $l^3$. The probability to find an electron inside a specific box is given by the sum over the squares of the corresponding wave amplitudes of the eigenstate $\phi$  
	\begin{align}
	\mu_b(\phi,l,L)=\sum\limits_{i\in \mathrm{box}~b}\left|\phi_i\right|^2\quad.
	\end{align}
	The mass exponent for moment $q$ is defined by 
	\begin{align}
	\tau_q(\phi,l,L)=\lim\limits_{\lambda\rightarrow 0}\frac{\ln P_q(\phi,l,L)}{\ln\lambda}\label{eqn:massexponent}
	\end{align}
	with the generalized mass
	\begin{align}
	 P_q(\phi,l,L)=\sum\limits_{b}\mu^q_b(\phi,l,L)
	\end{align}
	and relative box size $\lambda=l/L$. According to Eq.~(\ref{eqn:massexponent}) the value of $\tau_q$ can be extracted as slope from the linear regression of $\ln P$ versus $\ln\lambda$ with emphasis on small box sizes. In this work we compare several ranges between $0.1\leq\lambda\leq0.5$. 
	
	The singularity spectrum $f(\alpha_q)$~\cite{HalJKP86} contains further information about the wave function and can be calculated by a parametric representation of the singularity strength
	\begin{align}
	\alpha_q=\frac{\mathrm{d}\tau_q}{\mathrm{d}q}=\lim\limits_{\lambda\rightarrow 0}\frac{\sum_{b}\bar{\mu}^q_b\ln\mu_b}{\ln\lambda}\label{eqn:singularity_strength}
	\end{align}
	and the fractal dimension 
	\begin{align}
		f_q=q\alpha_q-\tau_q=\lim\limits_{\lambda\rightarrow 0}\frac{\sum_{b}\bar{\mu}^q_b\ln\bar{\mu}^q_b}{\ln\lambda}\label{eqn:fractal_dimension}
	\end{align}
	with
	\begin{align}
	\bar{\mu}_b^q=\frac{\mu_b^q}{P_q(\phi,l,L)}\quad.
	\end{align}
    Numerically $f_q$ and $\alpha_q$ are determined in the same way as $\tau_q$.
    
    In the standard box-counting algorithm the system is divided into boxes without overlap. Thus only box sizes with an integer ratio for $L/l$ are allowed. Depending on $L$ the number of possible $\lambda$ values can be very restricted. This limitation can be eliminated by using periodic boundary conditions together with a folding-back of protruding sites into the original system \cite{SchG91,ThiS13}.
    
    A finite single wave function is insufficient for a statistical statement. Therefore we use a set of wave functions and averaged quantities. Two different averages are commonly used, the ensemble (ens) and typical (typ) average~\cite{EveM08,VasRR08,RodVR08b}. Exemplified for the mass exponent one obtains
	    \begin{align}
	    	\tau_q^{\mathrm{ens}}=\lim\limits_{\lambda\rightarrow 0}\frac{\ln\left\langle P_q(\phi,l,L)\right\rangle_\phi}{\ln\lambda}\label{eqn:massexponent_ens}
	    \end{align}
	 and 
		 \begin{align}
			 \tau_q^{\mathrm{typ}}=\lim\limits_{\lambda\rightarrow 0}\frac{\left\langle\ln P_q(\phi,l,L)\right\rangle_\phi}{\ln\lambda}\quad \label{eqn:massexponent_typ}
		 \end{align}
    with arithmetic average $\langle\cdot\rangle_\phi$ over all considered wave functions. The averaged quantities from  Eqs.~(\ref{eqn:singularity_strength}) and (\ref{eqn:fractal_dimension}) can be derived analogously.
    
    For application of the MFA to the CST lattice, our system is stacked with primitive UCs and thus has a rhombohedral form. Therefore it is convenient to use rhombohedrons as boxes. The shape has no further effect on the algorithm. The primitive UC is used as the smallest possible box. A further subdivision seemed to be inappropriate, because empty boxes would follow.

 	\subsection{Green resolvent method}\label{sec:GRM}
	The GRM is based on a iterative formulation of the time-independent Green function $\mat{G}=(\mat{Z}-\mat{H})^{-1}$ for a system described by a Hamiltonian $\mat{H}$~\cite{Hay80,MacK81,MacK83}. $\mat{Z}=(E+\mathtt{i}\eta)\mat{I}$ is a diagonal energy matrix with the complex quantity $\eta$. 
	The system has a bar-shaped geometry with cross section  $L\times L$ and length $N\gg L$ and grows layer by layer at one end of the bar. This process is considered as a perturbation $\mat{\Delta}$ of the previous system $\mat{\tilde{H}}$ and thus expressed with the Dyson equation
	\begin{align}
	\mat{G}=\mat{\tilde{G}}+\mat{\tilde{G}}\mat{\Delta}\mat{G}\quad.\label{eq:Dyson}
	\end{align}
	$\mat{\tilde{G}}$ is the corresponding Green function of $\mat{\tilde{H}}$. The further description uses a notation where $\mat{G}_{i,j}^{(N)}$ is the submatrix of a system with length $N$ which couples the $i$th and $j$th layers. In the same manner the Hamiltonian $\mat{H}^{(N+1)}$ can be represented as tridiagonal block matrix
	with the layer Hamiltonians $\mat{H}^{(N+1)}_{i,i}$ as diagonal entries and the inter-layer connections $\mat{V}_{i,i+1}$ or $\mat{V}_{i+1,i}$ as nondiagonal entries. The matrix without the last row and column is the Hamiltonian $\mat{H}^{(N)}$ of the previous step. The last nondiagonal block matrices $\mat{V}_{N,N+1}$ and $\mat{V}_{N+1,N}$ are used for the perturbation $\mat{\Delta}$ of the matrix $\mat{\tilde{H}}^{(N+1)}$. $\mat{\tilde{H}}^{(N+1)}$ is the combination of the matrix $\mat{H}^{(N)}$ and the last diagonal entry $\mat{H}^{(N+1)}_{N+1,N+1}$. Now Eq.~(\ref{eq:Dyson}) can be simplified to the four cases
	\begin{align}
		&\mat{G}^{(N+1)}_{N+1,N+1}&=&~\mat{\tilde{G}}^{(N+1)}_{N+1,N+1}\left(\mat{I}+\mat{V}_{N+1,N}\mat{G}^{(N+1)}_{N,N+1}\right)\quad,\label{eqn:Gnn}\\
		&\mat{G}^{(N+1)}_{N+1,k\leq N}&=&~\mat{\tilde{G}}^{(N+1)}_{N+1,N+1}\mat{V}_{N+1,N}\mat{G}^{(N)}_{N,k}\quad,\label{eqn:Gn1}\\
		&\mat{G}^{(N+1)}_{j\leq N,N+1}&=&~\mat{G}^{(N)}_{j,N}\mat{V}_{N,N+1}\mat{G}^{(N+1)}_{N+1,N+1}\quad\text{, and}\label{eqn:G1n}\\
		&\mat{G}^{(N+1)}_{j\leq N,k\leq N}&=&~\mat{G}^{(N)}_{j,k}+\mat{G}^{(N)}_{j,N}\mat{V}_{N,N+1}\mat{G}^{(N+1)}_{N+1,k}\quad. \label{eqn:G11}
	\end{align}
	Combining Eqs. (\ref{eqn:Gnn}) and (\ref{eqn:G1n}) then yields
	\begin{align}
	\mat{G}^{(N+1)}_{N+1,N+1}=\left(\mat{Z}-\mat{H}_{N+1,N+1}-\mat{\Sigma}_N \right)^{-1}\label{eqn:Gnn_neu}
	\end{align}
	with $\mat{\Sigma}_N=\mat{V}_{N+1,N}\mat{G}^{(N)}_{N,N}\mat{V}_{N,N+1}$.
	For the analysis of the critical behavior we employ the  localization length $\Lambda$ of the wave function for a long bar-shaped geometry  \cite{JohK83a,NikM93}
	\begin{align}
		\begin{aligned}
			\frac{1}{\Lambda}&=-\frac{1}{2N}\lim\limits_{\eta\rightarrow 0}\lim\limits_{N\rightarrow\infty}\ln\tr\left|\mat{G}^{(N)}_{1,N}\right|^2\\
			&=-\frac{1}{N}\lim\limits_{\eta\rightarrow 0}\lim\limits_{N\rightarrow\infty}\ln\frobnorm{\mat{G}^{(N)}_{1,N}}\label{eqn:LL_Green}
		\end{aligned}
		\end{align}
	with Frobenius norm $\frobnorm{\cdot}$. The submatrix 
	\begin{align}
	\mat{G}^{(N)}_{1,N}=\mat{G}^{(1)}_{1,1}\prod_{i=1}^{N-1}V_{i,i+1}\mat{G}^{(i+1)}_{i+1,i+1}
	\end{align} can be calculated iteratively by using Eq.~(\ref{eqn:G1n}). 
	In order to stabilize this calculation numerically MacKinnon \textit{et al}.~\cite{MacK83} used a transformation on simple cubic lattices, which can be generalized by the following steps where the size $k$ reflects the number of layers comprised in one step:\\ i) The calculation is initialized by 
	\begin{align}
	\mat{B}_1=\mat{G}_{1,1}^{(1)}\prod_{j=1}^{k-1}V_{j,j+1}\mat{G}^{(j+1)}_{j+1,j+1}
	\end{align} with $b_1=\frobnorm{\mat{B}_1}$. $\overline{\mat{B}}_1=\frac{\mat{B}_1}{b_1}$ is the normalized matrix of $\mat{B}_1$. ii) At each further step we build
	\begin{align} 
	\mat{B}_{i+1}=\overline{\mat{B}}_i\prod_{j=1}^{k}V_{j,j+1}\mat{G}^{(j+1)}_{j+1,j+1}
	\end{align} and $\overline{\mat{B}}_{i+1}=\frac{\mat{B}_{i+1}}{b_{i+1}}$ with $b_{i+1}=\frobnorm{\mat{B}_{i+1}}$. The norms $b_i$ are related to the corresponding Green functions for an integer number of steps $N/k$ by
	\begin{align}
	\frobnorm{\mat{G}^{(N)}_{1,N}}=\prod_{i=1}^{N/k}b_i \quad .
	\end{align}
	Eq.~(\ref{eqn:LL_Green}) then transforms to
	\begin{align}
	\frac{1}{\Lambda}=-\frac{1}{N}\lim\limits_{\eta\rightarrow 0}\lim\limits_{N\rightarrow\infty}\sum_{i=1}^{N/k}\ln b_i=-\frac{1}{k}\lim\limits_{\eta\rightarrow 0}\lim\limits_{N\rightarrow\infty}\left\langle\ln b\right\rangle\quad.
	\end{align}
	Here $\left\langle\cdot\right\rangle$ describes the average over $N/k$ steps. Correspondingly, the statistical error of $\Lambda$ follows from
	\begin{align}
	\var\Lambda=\frac{\Lambda^4}{k^2}\var\left(\left\langle\ln b\right\rangle\right)\quad.
	\end{align}
	The step size $k$ is limited by the two following facts. On the one hand it should be small ($k\ll N$) with regard to the numerical stability. On the other hand it should be large ($k\gg 1$) to get an accurate error estimation. This is due to the fact that successive $b_i$ are correlated to each other \cite{GRM_Correlation}. That correlation will be reduced by increasing $k$.  We use $k=1000$. 
	
	\subsection{Energy-level spacing}\label{sec:ELS}
		\begin{figure}
			\resizebox{0.5\textwidth}{!}{%
				\includegraphics{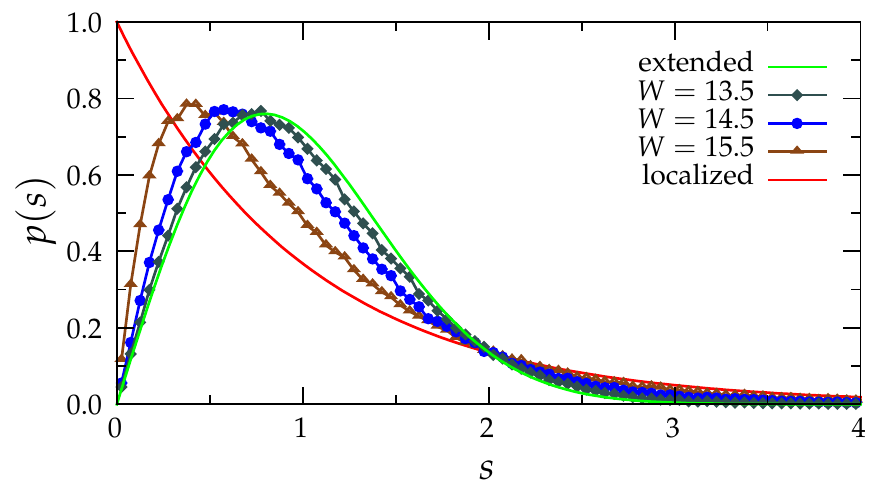}
			}
			\caption{Analytical level-spacing distributions for the extremely localized (Poisson distribution) and the extremely delocalized case (Wigner surmise) in comparison to numerical results for three disorder strengths near the transition with energy $E_\mathrm{c}=-4.0$. The system size is $L=50$.}
			\label{fig:LSD}
		\end{figure}
 	In contrast to the previous methods the ELS approach relies on the eigenenergies instead of the eigenstates of the system. We consider a subspectrum, which consists of the $N+1$ adjacent eigenvalues $\epsilon_k$ closest to energy $E$. The subspectrum is then unfolded by the normalized level spacings to obtain a dimensionless quantity 
	\begin{align}
		s'_i=\frac{\epsilon_{i+1}-\epsilon_i}{\left\langle\epsilon_{k+1}-\epsilon_k\right\rangle_k}
	\end{align}
	For the ELS distribution $p(s)=\left\langle\delta(s-s')\right\rangle_{s'}$ it follows that $\left\langle1\right\rangle_s =1$ and $\left\langle s\right\rangle_s=1$. The expected distributions are shown in Fig.~\ref{fig:LSD}.
	For the extremely localized case the eigenvalues are uncorrelated and the ELS distributions follow the Poisson law $p(s)^{(\mathrm{P})}=\exp(-s)$.
	In the delocalized case levels repel each other and the spacings are distributed according to the Wigner surmise $p(s)^{(\mathrm{W})}=\frac{\pi}{2} s\exp(-\frac{\pi}{4} s^2)$. All ELS distributions cross in a common point near $s_0=2.0$ which is utilized in the definition of a new measure \cite{HofS93}
	\begin{align}
	\gamma=\int\limits_{0}^{s_0}\mathrm{d}s\,p(s)=\left\langle\Theta(s_0-s')\right\rangle_{s'} \quad .
	\end{align}
 The average value of $\gamma$ over various realizations of the system for a given disorder strength $W$ allows us to increase the accuracy of our result and to estimate its statistical error by the standard deviation.
 
\section{Results}\label{sec:Results}
\begin{figure}
\resizebox{0.5\textwidth}{!}{%
	 \includegraphics{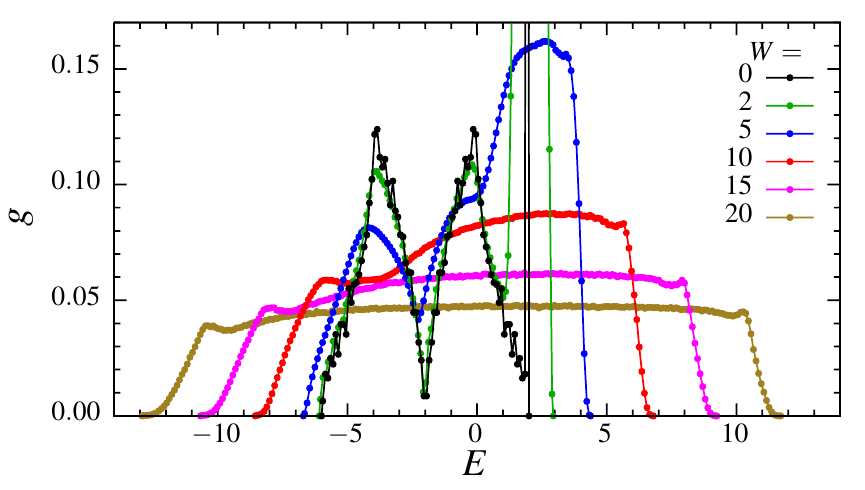}
	}
	\caption{DOS for a system of $19\times18\times17$ UCs for different disorder strengths $W$. The results are averaged over $200$ realizations of disorder. The bin width is $0.1$.}
	\label{fig:DOS}
	\end{figure}

	First we discuss the density of states (DOS). The DOS is calculated by diagonalization of the Anderson Hamiltonian\ (\ref{eqn:AH}). Its dependence on $W$ is shown in Fig.~\ref{fig:DOS}. Although the number of 6 nearest neighbors is the same as for the sc lattice, the DOS is very different. For systems without disorder ($W=0$) we obtain four separate energy bands due to the four basis sites. Two flat bands are degenerate and engender the delta peak at energy $E=2.0$, i.e. at the high energy band edge. The other two bands are distributed symmetrically with respect to $E=-2.0$. The total DOS spans the interval $E\in[-6,2]$ and is asymmetric in shape, which follows from the non-bipartiteness of the CST lattice. The energy range ensues from the fcc Bravais-lattice class and the number of nearest neighbors.  
	With increasing disorder strength $W$ the bands broaden as can be seen in Fig.~\ref{fig:DOS}. For $W=2$ the two dispersive lower bands are not very much affected, their DOS structures are only slightly washed out, while the two flat bands are strongly influenced and the delta peak is significantly widened to the interval $E\in \left[ 1,3\right]$ corresponding to the width of $W=2$ of the distribution of the on-site potentials.  The broadened delta peak quickly comprehends the energy band around $E=0$, reducing it to a shoulder in the DOS already for $W=5$. The minimum at $E=-2$ is then filled and for $W=10$ the lowest band is already a shoulder only. Finally, from the three maxima only small shoulders remain close to the tails of the distribution.
		
\begin{figure}
				\includegraphics{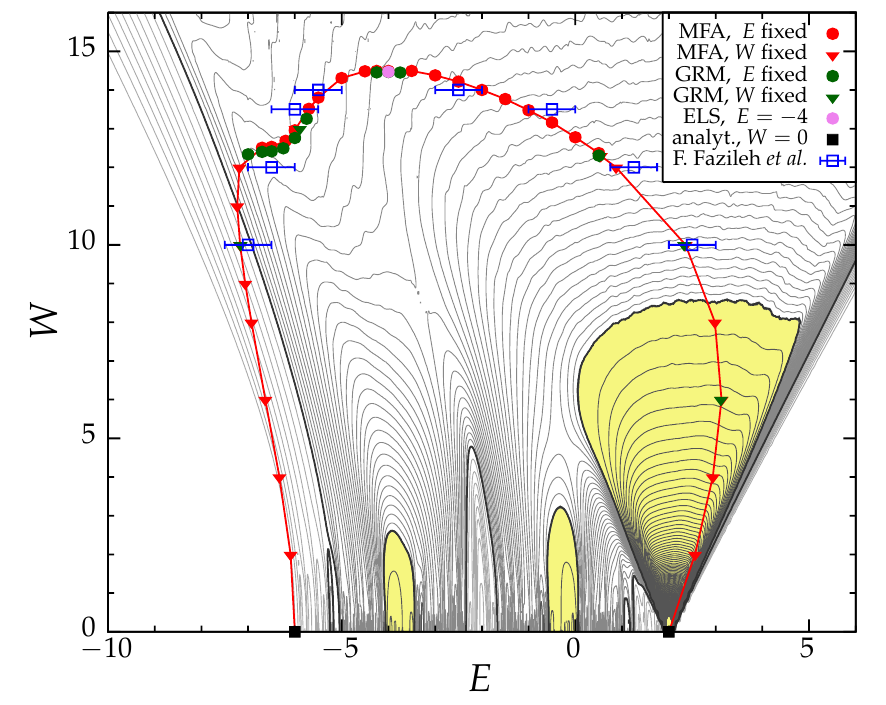}
				\caption{Phase diagram of the CST lattice. The area of extended states is enclosed by the red curve. Standard deviations for the data points are smaller than the red line width. The contour lines show the behavior of the DOS. This is divided into three areas (light, middle and dark grey lines; $\Delta g=0.004$, $\Delta g=0.01$ and $\Delta g=0.1$, respectively) which are separated by black curves at $g=0.04$ and $g=0.1$. Additionally, the regions of $g>0.1$ are shaded yellow.}
				\label{fig:phasediagram}
			\end{figure}

	We then investigate the $E$- and $W$-dependence of the mobility edge separating localized and delocalized states. 
	The corresponding phase diagram of the CST lattice is depicted in Fig.~\ref{fig:phasediagram}. 
	The critical points are mainly gathered from MFA and GRM. In the case of MFA we use system sizes from $L=20$ to $60$ with about $13$ points for each size in the interval $\rho\in[\rho_\mathrm{c}-1,\rho_\mathrm{c}+1]$ around the critical point. In the case of GRM we proceed alike. We use system sizes from $L=3$ up to $L=9$ and $N=4\cdot10^5$. In general, the irrelevant scaling behavior increases with decreasing $W$. This is more intense on the low energy side. Therefore critical points were calculated by GRM at higher disorder. With ELS we only calculated the transition at fixed energy $E=-4.0$, because of the high numerical effort. 
			\begin{figure*}
				\includegraphics{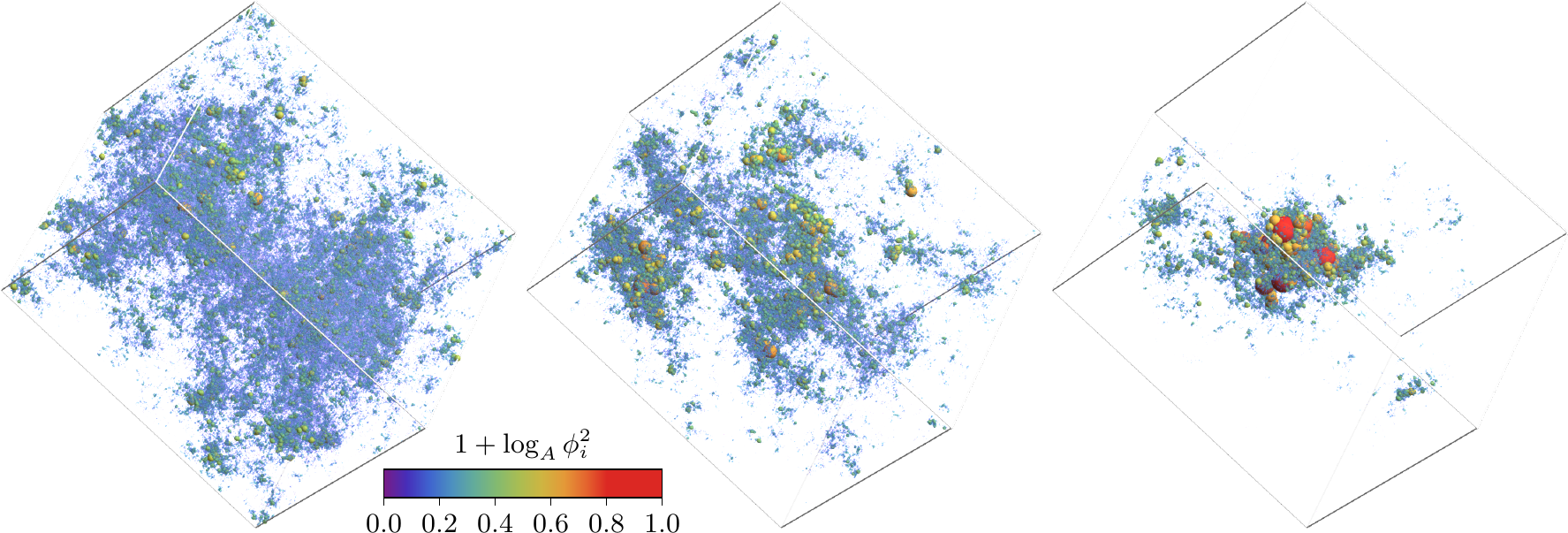}
				\caption{Eigenfunctions of a system with $120^3$ UCs at fixed energy $E=-4.0$ for three different disorder strengths for (from left to right) delocalized ($W=13.5$), critical ($W=14.5$), and localized ($W=15.5$) case. The on-site probability $|\phi_i|^2$ is visualized by the volume of the spheres and the indicated logarithmic color scale where $A$ denotes the number of sites. The darkest blue of the color scale would correspond to a homogeneously extended state, the brightest red to a state localized on a single site. }
				\label{fig:wavefunctions}
			\end{figure*} 
	In all cases the standard deviations of $E_\mathrm{c}$ and $W_\mathrm{c}$ are smaller than $0.02$. As a consequence of the shape of the DOS, also the mobility edge is asymmetric with respect to the mean energy. Our results resemble the findings of Fazileh \textit{et al.}~\cite{FazCGT06} but improve significantly on accuracy. The phase diagram bears some resemblance to that of the fcc lattice~\cite{EilFR08}. For low disorder strength the mobility edge follows the band edge. On the high energy side the mobility edge is quickly drawn into the region of very high DOS and then it tends towards the band center with increasing disorder, while on the low energy side the mobility edge enters the band tail slowly until it approaches the local maximum of the DOS. Here a shoulder-like structure appears in the mobility edge when the local minimum of the DOS is traversed. The largest critical disorder strength $W_\mathrm{c}=14.5$ at which all states are localized is observed at energy $E=-4.0$. We choose this transition to compare the results of MFA, GRM and ELS with respect to the universal behavior by high accuracy calculations.
	
\begin{figure}
		\includegraphics{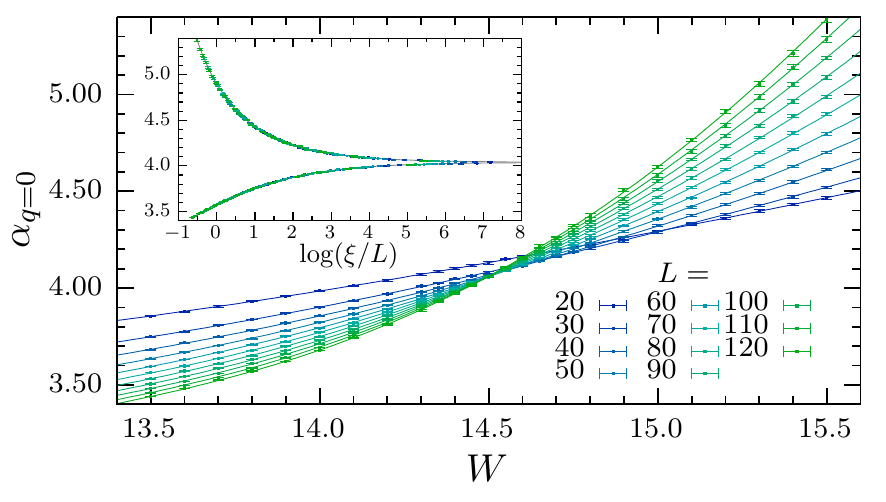}
		\caption{Behavior of $\alpha_{q=0}$ for the range $\lambda\in\left[0.1,0.4\right]$ of box sizes in dependence of disorder strength $W$ and system size. The lines are result of the fit with the expansion orders $n_\mathrm{r}=3$,  $n_\mathrm{i}=1$, $m_\mathrm{r}=2$, and $m_\mathrm{i}=0$ (see second line of Table~\ref{tab:results_mfa}). The inset displays the scaling function and the corrected data according to the fit.}
		\label{fig:LDT_mfa}
	\end{figure}

	For MFA we compute one state closest to energy $E=-4.0$ for each realization of disorder by diagonalization of the Anderson Hamiltonian (\ref{eqn:AH}) using a Jacobi-Davidson method with preconditioning \cite{BolN06}. Characteristic examples for localized, critical and extended wave functions are presented in Fig.~\ref{fig:wavefunctions}. For a statistical average a large number of realizations for each disorder strength $W$ is chosen in dependence on the system size $L$ as given in Table~\ref{tab:mfa_states}. 
		\begin{table}
				\caption{Number of wave functions used for each disorder strength depending on the number of UCs $L^3$ in the MFA.}
				\label{tab:mfa_states}
				\centering
				\begin{tabular}{c c c c c}
				\hline \rule[-2ex]{0pt}{5ex} $L$ & 20-60 & 70, 80 & 90, 100 & 110, 120 \\ 
				\hline \rule[-2ex]{0pt}{5ex} states & ${5000}$  & ${3000}$ & ${2000}$  &${1000}$ \\  
				\hline 
				\end{tabular}
		\end{table}
	The influence of the finite size effects can be controlled via the range of box sizes $\lambda$ used to fit the slope.
		\begin{table}
			\caption{Number of eigenvalues used for each disorder strength depending on the number of UCs $L^3$ in the ELS.}
			\label{tab:els_states}
			\centering
			\begin{tabular}{c c c c c}
				\hline \rule[-2ex]{0pt}{5ex} $L$ & 10-30 & 35, 40 & 45 & 50 \\ 
				\hline \rule[-2ex]{0pt}{5ex} eigenvalues & ${10000}$  & ${5000}$ & ${4500}$  &${2500}$ \\  
				\hline 
			\end{tabular}
		\end{table}
	 Leaving out small box sizes reduces the irrelevant contributions to FSS. In general a reduction of the number of box sizes increases uncertainties. We use the singularity strength $\alpha_{q=0}$ for the moment $q=0$ and compare different ranges of box sizes between $\lambda=0.1$ and $\lambda=0.5$. The results are given in Table~\ref{tab:results_mfa}. For $0.1\leq\lambda\leq0.4$ the irrelevant behavior is clearly evident and we obtain stable fits with $n_\mathrm{i}>0$ (see Fig.~\ref{fig:LDT_mfa}). The crossing point of the curves in Fig.~\ref{fig:LDT_mfa} yields the critical singularity strength $\alpha_{q=0}(W_\mathrm{c})\approx4.0$. This value is the same as in the case of Anderson localization in sc lattices~\cite{RodVSR11,RodVR08b} and vibrational localization in fcc lattices~\cite{LudTE03}. This supports the assumption that the multifractal spectrum at criticality depends only on the dimensionality and the Wigner-Dyson class~\cite{EveM08}.
	
\begin{figure}
		\includegraphics{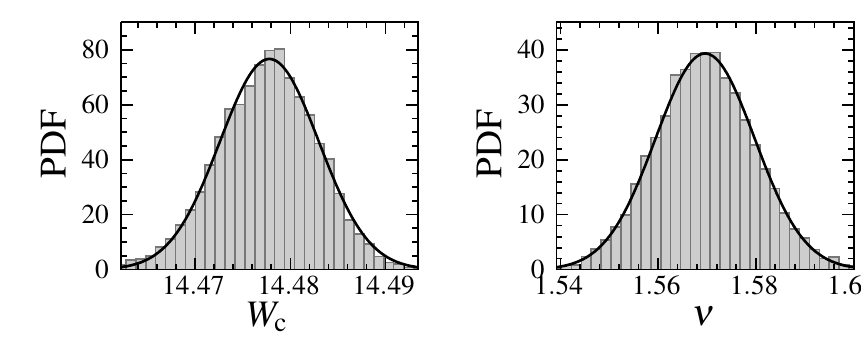}
		\caption{The histograms show the relative number of values (i.e. the probability distribution function) obtained for critical disorder strength $W_\mathrm{c}$ and critical relevant exponent $\nu$ by the error propagation calculation described in Sect.~\ref{sec:FSS} applied to the fit result in Fig.~\ref{fig:LDT_mfa}. The lines show the corresponding Gaussian curves.}
		\label{fig:LDT_mfa_error}
	\end{figure}

	The system-size dependent shift of the crossing point is nicely reproduced by the irrelevant part of the fit function (\ref{eq:fss}). The histograms of the corresponding error propagation in Fig.~\ref{fig:LDT_mfa_error} show Gaussian distributions.
	We therefore assume that the fit is stable and that we can use the standard deviations as measure for the statistical uncertainties. By increasing the irrelevant expansion order $n_\mathrm{i}$ or decreasing the irrelevant behavior it is more likely to get deviations from Gaussian distributions, e.g. the distributions are a superposition of several Gaussians. These results depend in particular on the initial variables of the minimization algorithm. For the range $0.3\leq\lambda\leq0.5$ the irrelevant behavior is significantly reduced. This is reinforced by omitting all values with $L<50$. These data can be represented by a scaling expansion without irrelevant exponent ($n_\mathrm{i}=0$). Such regressions are very stable and do not depend on the initial values strongly. In general, we find a higher relevant exponent $\nu$. For all results a single Gaussian distribution is observed. Despite the fact that these regressions simplify the actual behavior, they nevertheless confirm the results of the fits with higher irrelevant influences. 

	For MFA the quality $Q$ of fit is either very close to $1$ or very close to $0$ (for $0.1\leq\lambda\leq0.5$) indicating a problem in the error estimation of $\alpha_q$. The values of the errors can be overestimated in case of $Q=1$ or underestimated for $Q=0$.
	
	As a consistency check, we have also determined the probability density autocorrelation function
		\begin{align}
		C(R)=\frac{\left\langle|\phi_i|^2|\phi_j|^2\delta(|\vec{r}_i-\vec{r}_j|-R)\right\rangle_{i,j}-\left\langle|\phi_i|^2\right\rangle_{i}^2}{\left\langle|\phi_i|^4\right\rangle_{i}-\left\langle|\phi_i|^2\right\rangle_{i}^2}
		\end{align} 
	with the on-site probability $|\phi_i|^2$ at the real space position $\vec{r}_i$. In particular for $W=14.5$ the correlation function shows the expected power-law behavior $C(R)\propto R^{-c}$ in dependence on the Euclidean distance $R$. The exponent $c=1.51$ is close to the correlation dimension $D_{q=2}=\tau_{q=2}=1.39(2)$ obtained from Eq.~(\ref{eqn:massexponent}). As in Ref.~34 the deviation of $8\%$ can be attributed to finite size effects and the small number of samples. 

	For the calculation with GRM we build a long bar with  $N=4\cdot10^5$ UCs. The cross section $L\times L$ is always square shaped with system size $L$ ranging from $1$ to $12$ UCs. We use $50$ realizations of disorder for each size and the same disorder strengths as in the case of the MFA. The complex energy part $\eta$ is set to $10^{-6}$. The results for several expansion orders are listed in Table~\ref{tab:results_grm}. The irrelevant exponent $y\approx 8$ is higher than the value $y\approx 2$ from MFA.
	
	The eigenenergies required for ELS are computed by diagonalization of the Anderson Hamiltonian (\ref{eqn:AH}) analogously to determination of the eigenstates for MFA. We calculate the $101$ eigenenergies closest to a given energy $E$ for each realization of disorder. This method allows us to reach relatively large system sizes, it is, however, accompanied by a high numerical cost. With increasing system size we decrease the number of realizations (see Table~\ref{tab:els_states}). The fit results are given in Table~\ref{tab:results_els}.
	
\begin{figure}
			\includegraphics{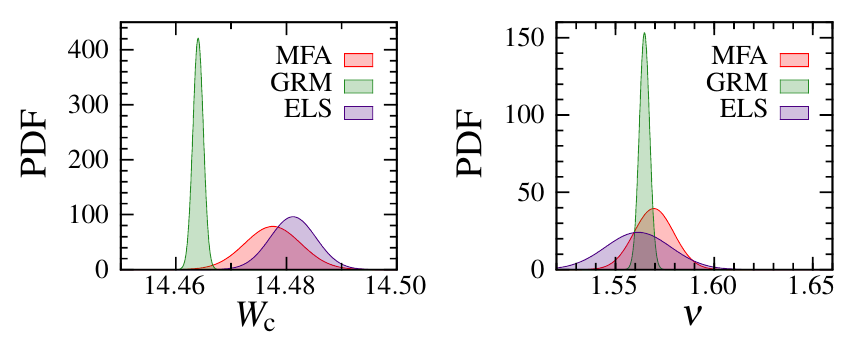}
			\caption{Probability distributions of the critical parameters for the transition at energy $E=-4.0$ are described by Gaussian distributions. The Gaussians of the MFA are shown in Fig. ~\ref{fig:LDT_mfa_error}. For the data of GRM and ELS see the first line of Table~\ref{tab:results_grm} and the seventh line of Table~\ref{tab:results_els}), respectively.}
			\label{fig:compare}
	\end{figure}

	The results of the error propagation for GRM and ELS are very similar to those of the MFA (Fig.~\ref{fig:LDT_mfa_error}). The results of all three methods are composed in Fig.~\ref{fig:compare}.
	The results and accuracy of MFA and ELS are very similar. GRM appears to be significantly more precise but agrees with the MFA and ELS results only in the value for the critical exponent $\nu$. For the critical disorder strength $W_\mathrm{c}$ the distributions do not overlap indicating an extra source of error not captured by our statistical analysis. Therefore, it appears not appropriate to use the error of the mean. Instead, we calculate the mean and variance of the sum of the three distributions. In this way the variance describes more than the pure random part. We obtain for the transition at energy $E_\mathrm{c}=-4.0$  the overall critical disorder strength $W_\mathrm{c}=14.474(8)$ and the universal critical exponent $\nu=1.565(11)$. While the value of $W_\mathrm{c}$ depends on the specific model the value of $\nu$ is a universal quantity and agrees well with recent publications \cite{RodVSR11,SleO14}.
	
	One possible source for the discrepancy of the $W_\mathrm{c}$ values in Fig.~\ref{fig:compare} could be that the investigated overall system sizes are qualitatively different: For MFA and ELS we consider large rhombohedrons (equivalent to large cubes) of the same extension in the three dimensions with periodic boundary conditions in all three directions. With GRM we treat quasi-one-dimensional bars of very large length (which explains the high accuracy reflected in the sharp distribution in Fig.~\ref{fig:compare}) but relatively small cross-section, i.e. small extensions in the other two dimensions. That could lead to enhanced interference effects due to periodic boundary conditions in these directions and thus to the somewhat stronger localization so that $W_\mathrm{c}$ is reduced.
	
 	\begin{table*}
 		\caption{The estimates of the critical parameters  $W_\mathrm{c}$, $\nu$, and $y$ with their standard deviations obtained from MFA for several expansion orders. The indices $\mathrm{l}$ and $\mathrm{u}$ denote the lower and upper bound of system size, respectively. $Q$ is the quality of fit. $N_\mathrm{P}$ and $N_\mathrm{F}$ are the number of points and degrees of freedom, respectively. For all transitions $\alpha^{\mathrm{ens}}_{q=0}$ was analyzed in the disorder range $W\in\left[13,5,15.5\right]$. We set $m_\mathrm{i}=0$ always.}\label{tab:results_mfa}
 		\centering
 		\begin{tabular}{|S[table-format = 2.4(2)]|S[table-format = 1.4(2),table-auto-round]|S[table-format = 1.4(4),table-auto-round]|S[table-column-width = 3mm, table-format = 1.1]S[table-column-width = 3mm, table-format = 1.1]|S[table-column-width = 3mm, table-format = 2]S[table-column-width = 3mm, table-format = 2]|S[table-column-width = 1mm]S[table-column-width = 1mm]S[table-column-width = 3mm]|S[table-column-width = 4mm, table-format = 3.0]|S[table-column-width = 4mm, table-format = 3.0]|S[table-format = 3.2,table-auto-round]|S[table-format = 1.4,table-auto-round]|}
 			\hline
 			{$W_\mathrm{c}$} &{$\nu$}  & {$y$} & {$\lambda_\mathrm{l}$} & {$\lambda_\mathrm{u}$} & {$L_\mathrm{l}$} & {$L_\mathrm{u}$}& {$n_\mathrm{r}$} & {$n_\mathrm{i}$} & {$m_\mathrm{r}$} & {$N_\mathrm{P}$} &{$N_\mathrm{F}$} & {$\chi^2$} & {$Q$}\\  
 			\hline
 			14,4775\pm0,0093 & 1,5694\pm0,0171 & 2,1560\pm0,1466 & 0,1 & 0.3 & 20 & 120 & 3 & 1 & 2 & 286 & 274 & 034,01 & 1,0000 \\
 			14,4775\pm0,0051 & 1,5694\pm0,0101 & 2,1598\pm0,1234 & 0,1 & 0.4 & 20 & 120 & 3 & 1 & 2 & 286 & 274 & 139,01 & 1,0000 \\
 			14,4757\pm0,0055 & 1,5653\pm0,0148 & 2,1432\pm0,1252 & 0,1 & 0.4 & 20 & 120 & 4 & 1 & 3 & 286 & 271 & 137,24 & 1,0000 \\
 			14,4609\pm0,0034 & 1,5889\pm0,0079 & 1,8628\pm0,1000 & 0,1 & 0.5 & 20 & 120 & 3 & 1 & 1 & 286 & 275 & 487,89 & 0,0000 \\
 			14,4747\pm0,0035 & 1,5654\pm0,0135 & 2,1172\pm0,1091 & 0,1 & 0.5 & 20 & 120 & 4 & 1 & 4 & 286 & 270 & 416,87 & 0,0000 \\
 			14,4705\pm0,0040 & 1,5892\pm0,0127 &                 & 0,3 & 0,5 & 50 & 120 & 3 & 0 & 1 & 208 & 202 & 99,41  & 1,0000 \\
 			14,4825\pm0,0046 & 1,5807\pm0,0127 &                 & 0,3 & 0,5 & 50 & 120 & 3 & 0 & 2 & 208 & 201 & 81,59  & 1,0000 \\
 			14,4796\pm0,0050 & 1,5906\pm0,0153 &                 & 0,3 & 0,5 & 50 & 120 & 3 & 0 & 3 & 208 & 200 & 80,05  & 1,0000 \\
 			14,4795\pm0,0051 & 1,5900\pm0,0158 &                 & 0,3 & 0,5 & 50 & 120 & 3 & 0 & 4 & 208 & 199 & 80,02  & 1,0000 \\
 			14,4692\pm0,0040 & 1,5788\pm0,0130 &                 & 0,3 & 0,5 & 50 & 120 & 4 & 0 & 1 & 208 & 201 & 91,13  & 1,0000 \\
 			14,4802\pm0,0054 & 1,5784\pm0,0132 &                 & 0,3 & 0,5 & 50 & 120 & 4 & 0 & 2 & 208 & 200 & 80,93  & 1,0000 \\
 			14,4805\pm0,0053 & 1,6015\pm0,0258 &                 & 0,3 & 0,5 & 50 & 120 & 4 & 0 & 3 & 208 & 199 & 79,79  & 1,0000 \\
 			14,4806\pm0,0056 & 1,6029\pm0,0290 &                 & 0,3 & 0,5 & 50 & 120 & 4 & 0 & 4 & 208 & 198 & 79,79  & 1,0000 \\
 			\hline
 		\end{tabular}
 	\end{table*}
	\begin{table*}
	\caption{Same as Table~\ref{tab:results_mfa}, but for the $\Lambda$ data from GRM.}\label{tab:results_grm}
	\centering
	\begin{tabular}{|S[table-format = 2.4(2)]|S[table-format = 1.4(2),table-auto-round]|S[table-format = 1.2(2),table-auto-round]|S[table-column-width = 3mm, table-format = 2]S[table-column-width = 3mm, table-format = 2]|S[table-column-width = 1mm]S[table-column-width = 1mm]S[table-column-width = 3mm]|S[table-column-width = 4mm, table-format = 3.0]|S[table-column-width = 4mm, table-format = 3.0]|S[table-format = 3.2,table-auto-round]|S[table-format = 1.4,table-auto-round]|}
	\hline
{$W_\mathrm{c}$} &{$\nu$}  &{$y$} & {$L_\mathrm{l}$} & {$L_\mathrm{u}$}& {$n_\mathrm{r}$} & {$n_\mathrm{i}$} & {$m_\mathrm{r}$} & {$N_\mathrm{P}$} &{$N_\mathrm{F}$} & {$\chi^2$} & {$Q$}\\  
	\hline
	14,4639\pm0,0008 & 1,5656\pm0,0019 & 8,28\pm0,26 & 1 & 12 & 4 & 2 & 2 & 300 & 281 & 337,09 & 0,0121 \\
	14,4640\pm0,0008 & 1,5634\pm0,0029 & 8,28\pm0,24 & 1 & 12 & 4 & 2 & 3 & 300 & 280 & 336,01 & 0,0121 \\
	14,4639\pm0,0008 & 1,5615\pm0,0031 & 8,33\pm0,25 & 1 & 12 & 4 & 2 & 4 & 300 & 279 & 333,11 & 0,0144 \\
	14,4640\pm0,0008 & 1,5627\pm0,0039 & 8,28\pm0,24 & 1 & 12 & 5 & 2 & 3 & 300 & 277 & 333,54 & 0,0112 \\
	14,4644\pm0,0008 & 1,5674\pm0,0017 & 8,23\pm0,23 & 2 & 12 & 3 & 1 & 2 & 275 & 263 & 328,16 & 0,0039 \\
	14,4644\pm0,0008 & 1,5673\pm0,0026 & 8,22\pm0,23 & 2 & 12 & 3 & 1 & 3 & 275 & 262 & 328,16 & 0,0034 \\
	14,4646\pm0,0009 & 1,5664\pm0,0024 & 5,20\pm0,20 & 2 & 12 & 3 & 2 & 2 & 275 & 259 & 312,23 & 0,0131 \\
	14,4645\pm0,0009 & 1,5676\pm0,0042 & 5,18\pm0,20 & 2 & 12 & 3 & 2 & 3 & 275 & 258 & 312,12 & 0,0118 \\
	14,4639\pm0,0008 & 1,5657\pm0,0019 & 8,27\pm0,25 & 2 & 12 & 4 & 1 & 2 & 275 & 261 & 320,63 & 0,0069 \\
	14,4640\pm0,0008 & 1,5632\pm0,0030 & 8,29\pm0,25 & 2 & 12 & 4 & 1 & 3 & 275 & 260 & 319,55 & 0,0069 \\
	14,4640\pm0,0009 & 1,5646\pm0,0026 & 5,33\pm0,28 & 2 & 12 & 4 & 2 & 2 & 275 & 256 & 307,27 & 0,0154 \\
	14,4640\pm0,0010 & 1,5633\pm0,0048 & 5,36\pm0,27 & 2 & 12 & 4 & 2 & 3 & 275 & 255 & 307,00 & 0,0142 \\
	14,4664\pm0,0010 & 1,5661\pm0,0022 & 8,71\pm0,31 & 2 & 10 & 3 & 1 & 2 & 225 & 213 & 245,95 & 0,0603 \\
	14,4665\pm0,0011 & 1,5658\pm0,0034 & 8,71\pm0,31 & 2 & 10 & 3 & 1 & 3 & 225 & 212 & 245,94 & 0,0549 \\
	14,4666\pm0,0013 & 1,5668\pm0,0055 & 5,49\pm0,36 & 2 & 10 & 3 & 2 & 3 & 225 & 208 & 236,52 & 0,0851 \\
	14,4658\pm0,0011 & 1,5643\pm0,0022 & 8,74\pm0,31 & 2 & 10 & 4 & 1 & 2 & 225 & 211 & 238,18 & 0,0964 \\
	14,4659\pm0,0011 & 1,5606\pm0,0037 & 8,76\pm0,32 & 2 & 10 & 4 & 1 & 3 & 225 & 210 & 236,86 & 0,0984 \\
	14,4657\pm0,0012 & 1,5626\pm0,0034 & 5,68\pm0,58 & 2 & 10 & 4 & 2 & 2 & 225 & 206 & 229,53 & 0,1250 \\
	14,4621\pm0,0011 & 1,5734\pm0,0045 &             & 5 & 12 & 3 & 0 & 3 & 200 & 192 & 232,48 & 0,0244 \\
	14,4618\pm0,0011 & 1,5689\pm0,0051 &             & 5 & 12 & 4 & 0 & 3 & 200 & 191 & 229,36 & 0,0302 \\
	14,4626\pm0,0010 & 1,5680\pm0,0026 &             & 5 & 12 & 3 & 0 & 2 & 200 & 193 & 234,62 & 0,0219 \\
	14,4619\pm0,0011 & 1,5665\pm0,0026 &             & 5 & 12 & 4 & 0 & 2 & 200 & 192 & 229,68 & 0,0326 \\
	\hline
	\end{tabular}
	\end{table*}
	\begin{table*}
	\caption{Same as Table~\ref{tab:results_mfa}, but for the $\gamma$ data from ELS. There are no irrelevant contributions taken into account, i.e. $n_\mathrm{i}=0$.}\label{tab:results_els}
	\centering
	\begin{tabular}{|S[table-format = 2.3(2)]|S[table-format = 1.4(2)]|S[table-column-width = 3mm, table-format = 2]S[table-column-width = 3mm, table-format = 2]|S[table-column-width = 1mm]S[table-column-width = 3mm]|S[table-column-width = 4mm, table-format = 3.0]|S[table-column-width = 4mm, table-format = 3.0]|S[table-format = 3.2,table-auto-round]|S[table-format = 1.4,table-auto-round]|}
	\hline
 {$W_\mathrm{c}$} &{$\nu$}  & {$L_\mathrm{l}$} & {$L_\mathrm{u}$}& {$n_\mathrm{r}$} & {$m_\mathrm{r}$} & {$N_\mathrm{P}$} &{$N_\mathrm{F}$} & {$\chi^2$} & {$Q$}\\ 
	\hline
	14,4754\pm0,0031 & 1,5453\pm0,0116                  & 10 &	50 & 3 & 1 & 234 & 228 & 232,33 & 0,4081 \\
	14,4755\pm0,0034 & 1,5456\pm0,0118                  & 10 &	50 & 3 & 2 & 234 & 227 & 232,30 & 0,3904 \\
	14,4761\pm0,0035 & 1,5697\pm0,0172                  & 10 &	50 & 3 & 3 & 234 & 226 & 227,37 & 0,4620 \\
	14,4771\pm0,0036 & 1,5674\pm0,0167                  & 10 &	50 & 3 & 4 & 234 & 225 & 225,49 & 0,4783 \\
	14,4755\pm0,0032 & 1,5438\pm0,0116                  & 10 &	50 & 4 & 1 & 234 & 227 & 230,54 & 0,4221 \\
	14,4825\pm0,0041 & 1,5490\pm0,0120                  & 10 &	50 & 4 & 2 & 234 & 226 & 223,22 & 0,5399 \\
 14,4812\pm0,0041 & 1,5614\pm0,0165                  & 10 &	50 & 4 & 3 & 234 & 225 & 221,90 & 0,5459 \\
	14,4814\pm0,0041 & 1,5613\pm0,0166                  & 10 &	50 & 4 & 4 & 234 & 224 & 221,83 & 0,5284 \\
 \hline
	\end{tabular}
	\end{table*}

   \section{Conclusion}\label{sec:Conclusion}
	We compared the scaling behavior at the LD transition obtained by different methods in order to analyze the consistency of the obtained critical parameters and the accordance with universality. A sophisticated error propagation analysis offers the basis to estimate the accuracy of our results. Generally, the differences between the methods are convincingly small. The averaged regular scaling exponent is $1.565(11)$ and emphasizes the universal character. This value is comparable with other results for the orthogonal universality class \cite{RodVSR11,SleO14,UjfV15}. However, the obtained critical disorder strengths do not agree completely, but the deviations are smaller than the disorder-sampling interval. The mean value is $W_\mathrm{c}=14.474(8)$. The discrepancy follows from unknown systematic deviations. The error estimates describe the random deviation only. Nevertheless, a thorough random error analysis is important to find an appropriate regression of the data. 
	
	For the phase diagram of the CST lattice we obtain a detailed shape of the mobility edge which is consistent with the results of Fazileh \textit{et al.} \cite{FazCGT06}. The shape of the mobility edge bears some resemblance to the fcc lattice~\cite{EilFR08}. We obtain a small shoulder on the lower energy side. The critical point at which all extended states disappear with increasing disorder is closer to the band center than in the fcc lattice. The sc, the bcc and the fcc lattice have the corresponding critical disorder strength $W_\mathrm{c}^\mathrm{sc}=16.53$, $W_\mathrm{c}^\mathrm{bcc}=20.81$, and $W_\mathrm{c}^\mathrm{fcc}=26.73$, respectively~\cite{RodVSR11,EilFR08}. As the LD transition is (at least in this energy range) an interference phenomenon, it is influenced by the number of nearest neighbor hopping terms $N$. For comparison, we use the relative critical disorder $w_\mathrm{c}= W_\mathrm{c}/N$. Applying this to the sc, the bcc, and the fcc lattice we obtain $w_\mathrm{c}^\mathrm{sc}=2.76$, $w_\mathrm{c}^\mathrm{bcc}=2.60$, and $w_\mathrm{c}^\mathrm{fcc}=2.23$, respectively. This suggests that the relative value decreases with increasing number of nearest neighbors. With $w_\mathrm{c}^\mathrm{CST}=2.41$ the CST lattice takes a value in this range close to the value of the fcc lattice although the number of nearest neighbors is the same as for the sc lattice. But as an interference phenomenon, the LD transition is also influenced by the return probability which in turn is influenced by the shortest length of closed loops of nearest neighbor connections, which is 3 in the CST lattice and thus the same as in the fcc lattice.    
	
	Apart from the results, the methods have shown advantages and disadvantages. In particular with GRM it is difficult to compute weakly disordered lattices because of the increasing irrelevant behavior that could be compensated only by higher numerical effort. On the other hand the measure $\Lambda$ of GRM can be calculated directly and therefore the calculation is very stable. In contrast, the other methods which are based on eigenvalues or eigenfunctions of large sparse matrices, are indirect and not as stable as GRM. The MFA can be used in a broad disorder range. Considering error propagation, the analysis of $\alpha_q$ by linear regression is problematic. The errors tend to be overestimated. Values of different box sizes may not be statistically independent. The ELS is simple to implement, but requires a large number of calculated eigenvalues or, to be more precise, lattice realizations. Consequently the system size is limited and thus restricts the accuracy of $\nu$ particularly.

\end{document}